# Solution to the Proton Radius Puzzle

D. Robson

*Dept. of Physics , Florida State University, Tallahassee, Florida, 32306, USA*

The relationship between the static electric form factor for the proton in the rest frame and the Sachs' electric form factor in the Breit momentum frame is used to provide a value for the difference in the mean squared charge radius of the proton evaluated in the two frames. Associating the muonic-hydrogen data analysis for the proton charge radius of 0.84087 fm with the rest frame and associating the electron scattering data with the Breit frame yields a prediction of 0.87944 fm for the proton radius in the relativistic frame. The most recent value deduced via electron scattering from the proton is 0.877(6) fm so that the frame dependence used here yields a plausible solution to the proton radius puzzle.

## I. INTRODUCTION

The proton radius "measurements" from muonic-hydrogen laser spectroscopy [1,2] and the recent elastic electron scattering data from the proton [3, 4] are known to be in serious disagreement by about 4%. The error in the latest muonic-hydrogen result of 0.84087 fm for the proton radius is only 0.00039 fm which is at 7.7$\sigma$ variance with respect to the most recent average proton radius value of 0.8772 fm from electron data [4] with an assigned error of 0.0046 fm . The most recent discussion [5] of this large disagreement between these two types of measurement assumes (as do most workers in this field of study) that the symbol $<r_p^2>$ is the same entity in each of the analyses of the two types of data. In fact it is assumed that the Sachs' electric form factor $G_E$ deduced from the electron scattering data is the *same* electric form factor "$G_E$" used to derive in [5] the leading term for the proton finite size energy shift $\Delta E$ of the atomic ns states in muonic-hydrogen, i.e. , using "$G_E(\mathbf{q}^2)$" = $1 - \mathbf{q}^2 <r_p^2>/6 +$…. corresponding to the conventional definition [5] of the mean squared charge radius $<r_p^2> = -6 d"G_E"/d\mathbf{q}^2$ evaluated at $\mathbf{q}^2 = \mathbf{q}\cdot\mathbf{q} = 0$ ; which yields

$$\Delta E = 2/3 \ \pi\alpha \ |\phi_{ns}(0)|^2 <r_p^2> . \qquad (1)$$

In (1) the lepton-proton non-relativistic wavefunction squared at the center-of mass of the proton is given by $|\phi_{ns}(0)|^2 = (\alpha\mu/n)^3/\pi$ in which $\alpha$ is the fine structure constant and $\mu$ is the usual lepton-proton reduced mass. A much more detailed and comprehensive derivation of (1) is given by Friar [6] that includes all finite size effects for light muonic atoms up to order $\alpha^6$. In [6] the quantity $<r_p^2>$ is shown to be the mean squared charged radius of the proton in terms of the proton (*at rest*) charge density $\rho_0(r_p)$, so that $<r_p^2> = \int r_p^2 \rho_0(r_p) d^3r_p$ and requires "$G_E(\mathbf{q}^2)$" = $\int \exp(i\mathbf{q}\cdot r_p) \rho_0(r_p) d^3r_p$ to be the 3D Fourier transform of the rest frame charge density. It is important to note that (1) is derived by Friar *without using "$G_E(\mathbf{q}^2)$".* In terms of perturbation theory the first order result for the proton finite size effect

(which arises from the difference between the proton finite size charge Coulomb potential and the proton point charge Coulomb potential) is given by <0|ΔV$_c$|0> in which ΔV$_c$ is *assumed* to be a local function of the coordinate r. Specifically ΔV$_c$ in [6] is given by :-

$$\Delta V_c(r) = -\alpha \int d^3s \, (1/|\mathbf{r}-\mathbf{s}| - 1/r) \rho_0(s) \,. \tag{2}$$

The matrix element <0|ΔVc|0> involves a specific atomic ns state labeled in coordinate space by <**r**|0> in [6]. Evaluating the matrix elements of ΔV$_c$ leads in [6] to a power series in $\alpha$ with the lowest power being $\alpha^4$ that yields identically the result in (1). The assumption that the modified Coulomb potential is a local function of the coordinate r is equivalent to assuming that the internal state of the composite proton in its' rest frame is Galilean invariant but not Lorentz invariant. *It is this local assumption that is central to our understanding of why the muonic hydrogen spectroscopy analysis yields a significantly smaller value for the proton mean square radius than the value found in the elastic electron scattering analysis.* This point is discussed in more detail in the next section. Further evidence for this point is provided using an example of a Lorentz invariant QCD –like quark model in section **III**.

In this report we focus on the difference between the most accurate atomic laser spectroscopy results and the most accurate electron scattering results. This means that the electron-hydrogen laser spectroscopy is not included here since its sensitivity to the proton radius is very much smaller than the muonic-hydrogen case due to the fact that the atomic states density at the origin scale as the lepton reduced mass to the third power. This is a sensitivity factor of 6.43 million in favor of the muonic case and accounts for the fact that even very precise measurements for hydrogen spectroscopy lead to very much larger errors than the muonic case. As pointed out by Pohl *et al* [5] half of the hydrogen spectroscopy measurements yield radii within 1$\sigma$ of 0.841 fm and the worst case involves only 3$\sigma$.

## II. MOVING FRAME DEPENDENCE OF ELECTRIC FORM FACTORS

Here the reason for the above difference in the proton radii is discussed in terms of the Licht-Pagnamenta (LP) theory [7,8] of the moving frame dependence of electric form factors for composite particles. For the proton in the muonic atom the proton and muon are very close to being in their rest frame but in electron scattering measurements the electron is very close to the relativistic limit of the velocity of light which in the Breit center of momentum frame has the proton initially moving with 3-momentum **p** and finally with 3-momentum −**p** corresponding to the 3-momentum transfer **q** being 2**p.** In the Breit frame the four momentum transfer is (0, **q**) and the Lorentz scalar invariant four momentum squared reduces to **q**$^2$. For the case of the electric form factors for the proton the relation given to relate the Breit frame form factor $G_{EpB}$ to the static rest frame form factor $S_p$ (="$G_{EpRF}$") given in the LP papers [7,8] is (assuming for simplicity point quarks):

$$G_{EpB}(\mathbf{q}^2) = (B(\mathbf{q}^2))^{(1-N)/2} \, S_p(\mathbf{q}^2/B(\mathbf{q}^2)) \,, \tag{3}$$

in which for the proton N=3 for the up, down quark content (uud) and

$$B(\mathbf{q}^2) = 1 + \mathbf{q}^2/(4M_p^2), \tag{4}$$

with $M_p$ in general being a function of $\mathbf{q}^2$ but normalized to the rest mass of the proton $M_0$ at $\mathbf{q}^2=0$. We have replaced the LP notation of $\alpha(\mathbf{q}^2)$ for the boost by $B(\mathbf{q}^2)$ to avoid conflict with the fine structure constant $\alpha$. The simple factor $B^{-1}$, which for N=3 appears as a boost factor multiplying the static rest frame function $S_p$ with the modified argument $\mathbf{q}^2/B$ in (3) above, arises from Lorentz contractions of the *internal* degrees of freedom in their direction of motion as is discussed in detail in the LP papers. The notation "$G_{EpRF}$" = $S_p$ is used here to indicate that the LP assumption for the proton rest frame electric form factor is *not* Lorentz invariant. However "$G_{EpRF}$" (= $S_p$) and the Sachs' Breit frame $G_{EpB}$ are the usual 3D Fourier transforms of their respective proton spatial charge densities. What is relevant here is the expansion of both form factors as a power series in $\mathbf{q}^2$ using the standard definition [5] for $<r_p^2>$ = $-6\, dG_{Ep}/d\mathbf{q}^2$ calculated at $\mathbf{q}^2 = 0$. This yields the relation between the proton mean squared radius evaluated in the rest frame ($<"r_p^2">_{RF}$) and in the Breit frame ($<r_p^2>_B$) as

$$<r_p^2>_B = <"r_p^2">_{RF} + 3/(2M_0^2), \tag{5}$$

in which "$G_{EpRF}$" is evaluated in the non-relativistic dynamics limit to yield $<"r_p^2">_{RF} = -6\, dS_p/d\mathbf{q}^2$ at $\mathbf{q}^2 = 0$. We note at this point that the leading finite size correction in the muonic-hydrogen analysis uses eq.(1) which assumes [6] that the proton and muon have an additive non-relativistic kinetic energy $\mathbf{p}\cdot\mathbf{p}/(2M_0) + \mathbf{p}\cdot\mathbf{p}/(2M_\mu) = \mathbf{p}\cdot\mathbf{p}/(2\mu)$ combined with a rest frame modified Coulomb potential. In Appendix F of [6] Friar states that in the non-relativistic approximation to lepton atomic states the effect of the motion of the nucleus (proton here) on the lepton Hamiltonian in the center-of-mass frame is simply the replacement of the lepton rest mass by the usual reduced mass. This is exactly equivalent to moving the atomic center-of mass to the proton center of mass which is then at rest. The use of the reduced mass $\mu$ in the muonic –atom finite size calculation is thus consistent with the use of the proton rest frame charge distribution in the definition of $<"r_p^2">$. The LP theory uses a proton rest frame static form factor $S_p$ that is consistent with the non-relativistic dynamics that dominate the finite size effects used in the muonic –hydrogen analyses.

The term $3/(2M_0^2)$ in (5) arises only from the boost factor $1/B$ which arises [7,8] from the Jacobian relating the Breit frame internal volume element to the rest frame volume element. This Jacobian for Breit frame elastic scattering form factors is the product of the spatial Lorentz contractions for each internal z-coordinate which is $M_p/(M_p^2+\mathbf{q}^2/4)^{1/2} = 1/B^{1/2}$. As discussed above for the proton for N=3 the boost is $1/B = 1- \mathbf{q}^2/(4M_p^2)+\ldots$ which yields from the definition of $<r_p^2>_B$ a contribution of $3/(2M_0^2)$. The fact that the mean squared radius may depend on the frame it is analyzed in is not discussed in the LP papers and appears to have been recognized only later in a paper by Stanley and Robson [9]. The modifications to the LP boost used in [9] were aimed at improving "relativized" quark model fits to the form factors at large $\mathbf{q}^2$ but those boost modifications to the LP theory do not apply at small $\mathbf{q}^2$. However the discussion in [9] (and references therein) points out that unitary transformations used to remove lower components from the Dirac quark states (so that only a Pauli spin representation remains) leads to an effective finite size for the quarks. More important is that such unitary transformations (as exemplified by the standard Foldy-Wouthuysen (FW) transformations) are non-local operators in radial-space [10]. In the muonic atom the modified Coulomb potential becomes non-local in the FW representation whereas in the Dirac 4-component theory for the quarks the modified Coulomb potential is a local function of the corresponding Dirac radial coordinate. Using a local modified Coulomb potential as in (2) above ignores the effect of averaging over the non-locality in r-space. The latter is due to removing the lower quark state components which exist to some extent in a relativistic description of the internal states of the composite proton. In effect the use of (2) in the analysis of the Lamb shift in muonic hydrogen corresponds to measuring the root mean square radius using only the non-relativistic or static components of the proton internal wavefunction.

It is important to understand that the term $3/(2M_0^2)$ in (5) is *not* the proton Darwin-Foldy (DF) term. As was discussed by Friar, Martorell and Sprung [11] in 1997 and more recently in 2011 by Jentschura [12] the Darwin-Foldy term is not contained in the Sachs' electric form factor which is the one photon exchange matrix element in first order Born approximation. Confusion arises because the DF term contributes $3/(4M_0^2)$ to the proton mean square radius if it is included in the radius squared definition via the use of a modified Sachs' form factor given by $G_{Ep}/(1 + \tau)^{1/2}$ with $\tau = \bm{q}^2/(4M_0^2)$. The fact is that only the conventional Sachs' form factors are used for the neutron and proton to define their mean square radii. The LP theory and both the muonic-hydrogen and elastic electron scattering analyses are all consistent in using the appropriate electric form factors in each frame to define the proton mean square radius as $-6\, dG_{EpB}/d\bm{q}^2$ and $-6\, dS_p/d\bm{q}^2$ at $\bm{q}^2 = 0$. The Sachs' electric form factor extracted from elastic electron scattering involves the cross section after radiative and two photon exchange corrections have been made. Similar corrections are also made in the muonic–hydrogen analyses. The most up to date theory of the 2S-2P Lamb shift splitting is given by A. Antognini *et al* in [13] which includes an improved description of the third Zemach moment contribution using a consistent quantum field framework for two-photon exchange diagrams which include the finite size Zemach part. The term given in (1) here is listed in [13] as being 99.5% of all terms involving $<r_p^2>$ of which the majority of the 0.5% terms arises from finite size effects on one- loop e-vacuum polarization contributions. All other finite size effects (such as $<r_p^k>$ with k >2) are too small to affect the extracted value of $<r_p^2>$.

### III. MEAN SQUARE RADIUS DIFFERENCE

Using the muonic -hydrogen measurement of $<"r_p^2"> = (0.84087(39))^2$ fm$^2$ for the rest frame value in (5) and adding the boost term ( 0.0663448 fm$^2$ ) yields the Breit frame result that we associate with elastic electron scattering measurements as $(0.87944(37))^2$fm$^2$.

It is important to reiterate that the leading finite size effect in eq.(1) above is given using Friar's assumption of a simple modified Coulomb potential which is interpreted here as a non-relativistic "model" for the proton internal structure when the proton is at rest. Similarly the proton internal structure (finite size) derived from electron scattering is assumed to involve a Lorentz invariant relativistic model for the proton internal structure derived from the infinite momentum Breit frame. To better understand the source of the difference between the two measurements a QCD - based relativistic quark model discussed by Abe and Fujita [14] is now discussed. Their model describes the internal state of the proton in terms of a massless Dirac equation with a Lorentz scalar linear confinement potential for each of the three quarks. The Dirac equation in [14] is given in terms of the usual Dirac matrices $\alpha$ and $\beta$ as :-

$$[\alpha.p + \beta\mu r]\Psi = W\Psi,$$

where $\mu$ is the strength of the potential. The squared Dirac equation can then be written as

$$[p.p + i\mu\beta\alpha_r + \mu^2 r^2]\Psi = W^2\Psi, \qquad (6)$$

In which $\alpha_r = \alpha.r/r$. Initially here for the proton we use the accurate approximate solutions of (6) suggested by Abe and Fujita. The large and small components of $\Psi$ are denoted in [14] by $G(r)\chi_{\kappa,\nu}$ and $-iF(r)\chi_{-\kappa,\nu}$ respectively with $\kappa = -1 = -(j+1/2)$ for the proton corresponding to L=0 for the large component spherical spinor and L=1 for the small component spherical spinor. The approximation used by Abe and Fujita is to replace G(r) and F(r) by the lowest $W^2$ oscillator solutions of the operator $p.p + \mu^2 r^2$ and specifically these are $G(r) \approx AR_{00}(r)$ and $F(r) \approx BR_{01}(r)$ in which $R_{nL}(r)$ are radial harmonic oscillator solutions with an oscillator length $a = 1/\mu^{1/2}$. The energy squared of each of these two components are $3\mu$ and $5\mu$. To find the approximate value of $W^2$ using the full operator in (6) above involves solving coupled equations since the term $i\mu\beta\alpha_r$ connects the large and small components. The results are analytic since A and B are determined by the equations : $[3\mu - W^2]A = \mu N_{01}B$, and $[5\mu - W^2]B = \mu N_{01}A$ with $N_{01}$ being the radial overlap integral over $R_{00}(r)$ and $R_{01}(r)$. For the proton one obtains the approximate value of $W^2$ as $[4 - (1+N_{01}^2)^{1/2}]\mu$ and the approximate value of the ratio $B^2/A^2$ is independent of $\mu$ and takes a specific value of 0.152441. This follows from the analytic value of $N_{01} = \{8/(3\pi)\}^{1/2}$ which is given approximately as 0.92 in Table 1 in [14]. Also the normalized Dirac solution satisfies $A^2 + B^2 = 1$ so that the approximate solution from [14] has $B^2 = 0.132277$ *independently* of the value of $\mu$. The mass of the proton ($M_0$) in the three (uud) quark model is $3W$ without center of mass correction and from [14] the accurate estimate of the value of $M_0$ with c.m. correction is $(6)^{1/2}W$.

The item of interest here is the mean squared charge radius of the proton which is given by the matrix element of the operator $\Sigma e_i r_i^2$, which for all quarks in the same spatial state and with the u ($e_u=2/3$ e) and d ($e_d=-1/3$e) quarks being an isospin doublet reduces to matrix elements of $r^2$ (= $r_i^2$ for all i) in radial space. The matrix elements for the approximate solutions have the values of 3/2 ($A^2/\mu$) and 5/2 ($B^2/\mu$) since $1/\mu = a^2$. Noting that if the solutions are corrected to the center of mass corresponds to replacing $<r^2>$ by 2/3 $<r^2>$ then we obtain (using the approximate model) the value of the mean squared charge radius as

$$<r^2> = (A^2 + B^2)/\mu + (2/3)B^2/\mu = 1/\mu + (2/3)B^2/\mu. \qquad (7)$$

The first term $1/\mu$ in (7) is the mean squared radius when B =0 and A=1 and occurs if the quark motions in the c.m. are described by a two component Pauli spin wave function as is used in the old non-relativistic model. Choosing $\mu^{1/2} =1/a = $ 0.23467 GeV or a = 4.2613 GeV$^{-1}$ in natural units so that in conventional size units a = 0.84087 fm. This term is the one we associate with the muonic-hydrogen radius measurement since as discussed above it is the static limit model that arises in the Friar [6] approach to the leading finite size effect used in the Lamb shift analyses. The additional term with $B^2 = 0.132277$ has the value 0.062352 fm$^2$ and yields the rms charge radius for the four component quark motion in the c.m. as 0.8772 fm. This radius is the one to be associated with the Breit frame value derived in the elastic electron scattering measurements. This value of $\mu$ yields $W$ = 0.38113 GeV, and that yields $M_0 = (6)^{1/2}W$ = 0.93403 GeV in close agreement (99.5%) with the known rest mass of the proton (0.93827 GeV). One can clearly see that the simple model is in very close agreement with the results given via the LP approach [7,8]. In particular the term $(2/3)B^2a^2$ = 0.062352 fm$^2$ is in good agreement with the "boost" term value of 0.066345 fm$^2$. More accurate calculations using additional radial oscillators and a small "current mass" for the quarks yield very accurate agreement with the LP results and for $M_0$. Since this is a relativistic quark *model* the use here is only intended to provide evidence for the possible importance of the small components of the Dirac quark solutions which are ignored completely in the muonic analyses.

## IV. CONCLUSIONS

This value of 0.8794 fm derived using (5) for the proton radius in the Breit frame is well within the assigned error of .0060 fm for the most recent [15] electron scattering value of 0.8770 fm. The effect of the boost term appears to very accurately remove the large disagreement between the two types of measurement and suggests that both types of measurement are complementary, although the muonic-hydrogen experiments are at least an order of magnitude more accurate. The accuracy of the LP theory for small $\mathbf{q}^2$ depends upon the details of the internal wavefunctions used to describe the proton at rest. The three valence quarks' motion can be close to the infrared limit of QCD as discussed by this author [16] when local SU(3) color gauge invariance is imposed.

The existence of sea quark effects (e.g., pionic cloud effects) is not accurately known for the proton (nor the neutron). In the quark model approach by Geiger and Isgur [17] they find that there are strong cancellations between the hadronic components of the q-qbar sea which tend to make it transparent to photons. At the same time as [17] the work of Buchmann, Hernandez and Faessler [18] suggested that the charged $\pi$, gluonic and confinement exchange current effects are significant particularly for the neutron charge radius. However the confinement terms used in [18] are between pairs of quarks in violation of local SU(3) color gauge invariance as discussed in [16]. It is unclear therefore if the results in [18] are a reliable measure of sea quark effects on nucleon radii. A useful later discussion in (1999) of the rest frame charge distribution and $G_{En}$ for the neutron has been given for the valence quark model by Isgur [19]. Note however that the neutron mean square radius is *unaffected independently of the model* by the boost terms (as is the case also for the $\pi^0$ or any zero charged hadron) because the leading term in the chargeless hadron rest frame form factor is already of order $\mathbf{q}^2$ so that the boost terms only affect terms of higher order than $\mathbf{q}^2$. Because of this result for the neutron the inclusion of small components of pionic clouds (such as the $\pi^+$ - n configuration) has no significant effect on the proton radius. This occurs because the boost term at order $\mathbf{q}^2$ for this pionic configuration is the same as the uud proton component since it also involves only two internal coordinates contributing to the boost at order $\mathbf{q}^2$. These two coordinates are the internal u-dbar separation in the $\pi^+$ and the relative coordinate between the $\pi^+$ and the neutron.

The prediction from the major result above is that more accurate hydrogen spectroscopy measurements of $<r_p^2>$ in the future will obtain values more closely in accord with the muonic-hydrogen values. The suggestion of a potential systematic error in the available hydrogen spectroscopy data has been presented already [20].

## ACKNOWLEDGEDGEMENTS

I am grateful to Jerry Miller and Thomas Walcher for many enlightening discussions of the proton radius puzzle.